\documentclass[onecolumn,10pt,draftcls]{IEEEtran}

\usepackage{cite}
\usepackage{tikz}
\usepackage{blindtext}
\usepackage{graphicx}
\usepackage{amsmath}
\usepackage{relsize}
\usepackage{mathrsfs}
\usepackage[capitalise]{cleveref}
\usepackage[sans]{dsfont}
\usepackage{amsbsy}
\usepackage{amssymb}
\usepackage{lipsum}
\usepackage{enumerate}
\usepackage{float}
\usepackage{algorithm}
\usepackage{algpseudocode}
\usepackage{calc}
\usepackage{algorithmicx}

\usetikzlibrary{patterns}

\DeclareMathOperator{\Equaldef}{\overset{def}{=}}
\renewcommand{\Pr}{\mathbb{P}}

\newtheorem{problem}{Problem}
\newtheorem{proposition}{Proposition}
\newtheorem{remark}{Remark}
\newtheorem{theorem}{Theorem}
\newtheorem{definition}{Definition}
\newtheorem{corollary}{Corollary}

\title{Data-driven sensor scheduling for remote estimation in wireless networks}
\author{Marcos M. Vasconcelos and Urbashi Mitra\thanks{M. M. Vasconcelos and U. Mitra are with the Department of Electrical Engineering, University of Southern California, Los Angeles, CA 90089 USA. E-mails: \texttt{\{mvasconc,ubli\}@usc.edu}.}}

\begin{document}

\maketitle

\begin{abstract}
Sensor scheduling is a well studied problem in signal processing and control with numerous applications. Despite its successful history, most of the related literature assumes the knowledge of the underlying probabilistic model of the sensor measurements such as the correlation structure or the entire joint probability density function. Herein, a framework for sensor scheduling for remote estimation is introduced in which the system design and the scheduling decisions are based solely on observed data. Unicast and broadcast networks and corresponding receivers are considered. In both cases, the empirical risk minimization can be posed as a difference-of-convex optimization problem and locally optimal solutions are obtained efficiently by applying the convex-concave procedure. Our results are independent of the data's probability density function, correlation structure and the number of sensors.
\end{abstract}

\section{Introduction}
Sensor scheduling is a classical problem in signal processing and control with a very rich history. The traditional static sensor scheduling problem consists of selecting a subset of $k$ sensors among a group of $n$ sensors such that the expected distortion between the random state-of-the-world and its estimate is minimized \cite{Moon:2017}. The fact that we are selecting $k$ out of $n$ sensors already indicates that this problem is of combinatorial nature and typically hard to solve. This class of problems has many applications in engineering, especially in sensor networks in which the number of sensors allowed to communicate with a remote fusion center is limited due to bandwidth constraints. In an extreme case, the sensor scheduling problem consists of choosing one among $n$ sensors, and transmitting only its measurement across the network.

Consider the system described in the block diagram of \cref{fig:diagram}, where $n$ sensor-estimator pairs share a common wireless network, which can operate either in unicast or broadcast modes. 
The system operates as follows. Each of the $n$ sensors observe a distinct random variable and reports it to the scheduler. The scheduler selects a single random variable according to a scheduling decision rule and transmits it over the network. If the network is in unicast mode, only the intended estimator receives the sensor's observation and the remaining estimators observe an erasure symbol. If the network is in broadcast mode, all the sensors receive the same transmitted measurement. Upon observing the network output, each receiver forms its estimate according to an estimation policy. The goal of the system designer is to select scheduling and estimation policies such that the mean-squared error (MSE) between the observations at the sensors and the estimates at the receivers is minimized. This problem lies in the category of team decision problems with non-classical information structure, which are in general very difficult to solve due to a coupling between the scheduling and estimation policies known as signaling \cite{Ho:1980}. 

In addition to the classical applications of sensor scheduling, the framework proposed here can be used to model real-time communication between Internet of Things (IoT) devices at the edge. Due to the massive number of devices and the very high demand for communication resources, the scheduler selects the pieces of information that are most relevant for a given task and discard the others, keeping the network data flow under control but at the same time achieving a good system performance. A more specific application of interest is in systems known as Wireless Body Area Networks (WBANs) for remote health care monitoring \cite{Mitra:2012,Zois:2013,Zois:2016}. In these systems, the sensors collect heterogeneous biometric data and transmit them to a mobile phone, which acts as a scheduler. In order to preserve battery life and meet bandwidth constraints, the mobile phone selects one of them to transmit over the network to one or multiple destinations. 

To the best of our knowledge, most of the literature in sensor scheduling assumes that the joint probability density function (PDF) of the random variables observed at the sensors is known \emph{a priori} to the system designer. However, this is a restrictive assumption because in most practical applications this information is typically not available. The main challenge that we are trying to address in this paper is to design such system in the absence of knowledge of the joint PDF, but in the presence of a dataset of independent and identically distributed (IID) samples. As it is to be expected in such situations, very few theoretical guarantees can be provided under this general set of assumptions. The set of theoretical results and algorithms presented here are based on ideas from quantization theory \cite{Gray:2006} and modern techniques in non-convex optimization theory \cite{Shen:2016}. The results herein are mean to be provide a guide to the art of designing such complex data-driven scheduling for remote estimation systems.

The main contributions of this work are:

\begin{itemize}
	\item We provide a systematic data-driven approach for the joint design of scheduling and estimation rules for unicast and broadcast networks.

	\item Our algorithm exploits decompositions of non-convex objectives as a difference-of-convex (DoC) functions and use the convex-concave procedure (CCP) to efficiently find locally optimal solutions with guaranteed fast convergence.

	\item Our algorithms are universal and work irrespective of the joint PDF that generated the dataset and for any number of sensors.

	\item We establish a connection between our algorithms and subgradient methods. The main advantage of our algorithms is that we do not need to select a step size at every iteration in a \emph{ad hoc} manner. Our step sizes are constant and arise naturally from the CCP.
\end{itemize}

\subsection{Related Literature}

Sensor scheduling for remote estimation has a vast and ever growing literature dating back to the 1970's with the pioneering work of \cite{Athans:1972}. These problems are difficult in general, but certain formulations admit optimal solutions under assumptions on the underlying probabilistic model of the observations \cite{Shi:2012,Moon:2017,Vasconcelos:2019c}. The modern literature on this topic has addressed a number of issues ranging from energy constraints \cite{Mo:2011}, design of optimal event-triggered scheduling policies \cite{Weerakkody:2016}, energy-harvesting sensors \cite{Nayyar:2013,Michelusi:2013a},  strategic communication \cite{Farokhi:2017,Gao:2018}, and performance-complexity trade-offs \cite{Zare:2018}.

Our problem is based on the Observation-Driven Sensor Scheduling (ODSS) framework introduced in \cite{Vasconcelos:2019b}, where the scheduling of sensors making correlated Gaussian observations is considered. The work in \cite{Vasconcelos:2019b} uses team decision theory to obtain person-by-person optimal scheduling and estimation policies while seeking to prove the optimality of the so-called \emph{max-scheduler} proposed in \cite{Xia:2017}, which consists of letting the sensor with the measurement of largest magnitude transmit over the network. The subsequent work \cite{Vasconcelos:2019c} considered a sequential ODSS framework with an energy-harvesting scheduler for sensors making independent observations distributed according to symmetric and unimodal PDFs. 

In this work, we study the ODSS framework under minimal assumptions on the probabilistic model. Our goal is to design systems that: 1. could be used for any joint PDF without assumptions on the correlation structure of the sensor observations; 2. provide a learning framework that could guide the designer in choosing a scheduler with performance likely to be close to the optimal in case the PDF is unavailable. Our approach follows the current state of the art in learning for controls and estimation, where models are not fully available to the system designer and must be learned from data \cite{Baggio:2019,Battistelli:2016,Liu:2017}.
The DoC decompositions of the objective function contained here were observed in \cite{Vasconcelos:2019b} for the scheduling of two sensors making correlated Gaussian observations. Here we formally establish the results in full generality and in addition, we show the connection of the resulting CCP with subgradient methods with constant step sizes. 
Unfortunately, due to the lack of convexity and without knowledge on the probabilistic model, we cannot guarantee that the solutions found by our algorithms are in fact optimal, but we provide a learning framework which provides a systematic way to train and validate the performance of the data-driven design.

\subsection{Notation}
We adopt the following notation: random variables and random vectors are represented using upper case letters, such as $X$. Realizations of random variables and random vectors are represented by the corresponding lower case letter, such as $x$. 
The probability density function of a continuous random variable $X$ is denoted by $f_X$. 
 The real line is denoted by $\mathbb{R}$.
 The probability of an event $\mathfrak{E}$ is denoted by $\Pr(\mathfrak{E})$; the expectation of a random variable $Z$ is denoted by $\mathbb{E}[Z]$. The indicator function of a statement $\mathfrak{S}$ is defined as follows: 
\begin{equation}
\mathbb{I}\big(\mathfrak{S}\big) \Equaldef \begin{cases}
1 & \text{if} \ \ \mathfrak{S}\ \  \text{is true}\\
0 & \text{otherwise}.
\end{cases}
\end{equation}

\begin{figure}[ht!]
\begin{center}
\includegraphics[width=0.45\textwidth]{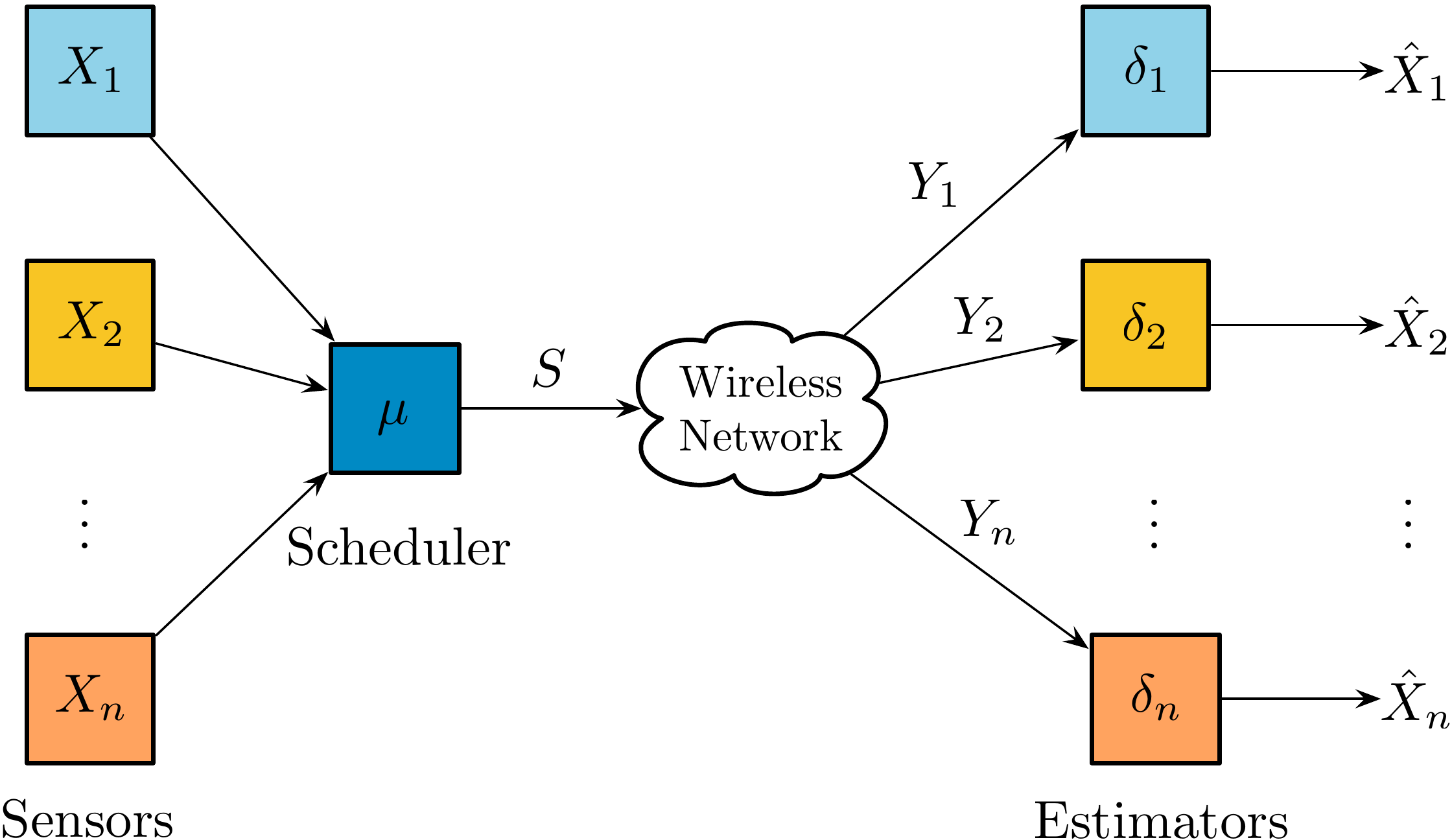}  
\caption{Schematic diagram for the remote sensing system with $n$ sensor-estimator pairs over a bandwidth constrained wireless network.} 
\label{fig:diagram}
\end{center}                                
\end{figure}

\section{Problem formulation}

Consider the system depicted in \cref{fig:diagram} with $n\geq 2$ sensor-estimator pairs communicating via a constrained wireless network.  We assume that the data observed at the sensors are realizations of the following continuous random vector 
\begin{equation}
X\Equaldef (X_1,X_2,\cdots, X_n),
\end{equation} 
which is  distributed according to an arbitrary joint PDF, $f_X$. We also assume that each $X_i$, $i\in\{1,\cdots,n\}$ has finite first and second-order moments. These are the only assumptions on the underlying probabilistic model of the problem.

The measurements are communicated by the sensors to a scheduler. Due to bandwidth constraints, we assume that only one sensor measurement can be transmitted at a time and the role of the scheduler is to choose which of the sensor measurements is transmitted over the network to its destination. The scheduling decision, $U\in\{1,\cdots,n\}$ is taken according to a policy $\mu: \mathbb{R}^n \rightarrow \{1,\cdots,n\}$ such that
\begin{equation}
U = \mu(X_1,\cdots,X^n).
\end{equation}
When a sensor is chosen by the scheduler, a communication packet $S$ containing its measurement and identification number is sent over the network, \emph{i.e.}, if $U=j$ then
\begin{equation}
S = (j,X_j).
\end{equation}

In this work, we will consider \textit{unicast} and \textit{broadcast} networks. In the case of a \underline{unicast network}, only the estimator associated with the chosen sensor receives the transmitted measurement. The remaining estimators receive a special erasure symbol $\varnothing$. In other words, if $U=j$ then 
\begin{equation}
Y_i = \begin{cases}
(j,X_j), & i=j \\
\varnothing,  &  i\neq j.
\end{cases}
\end{equation}
When the scheduling policy is properly designed, the erasure symbol may convey valuable information about $X_i$ to its corresponding estimator. In the case of a \underline{broadcast network}, whichever packet transmitted by the scheduler is received by all the estimators, \emph{i.e.}, if $U=j$ then
\begin{equation}
Y_i = (j,X_j) \ \ i\in\{1,\cdots,n\}.
\end{equation}

Upon receiving $Y_i$, the $i$-th estimator uses a function $\delta_i$ to compute an estimate of the $i$-th measurement as follows
\begin{equation}
\hat{X}_i = \delta_i(Y_i), \ \ i\in\{1,\cdots,n\}.
\end{equation}
We denote the collection of estimation functions by 
\begin{equation}
\delta \Equaldef (\delta_1,\cdots,\delta_n). 
\end{equation}

\begin{problem}[Observation-driven sensor scheduling]\label{prob:main}
Given the joint PDF of the sensor data $f_X$ and the network operation mode (unicast or broadcast), design the scheduling and estimation policies $\mu$ and $\delta$ such that the following mean-squared error between observations and estimates:
\begin{equation}\label{eq:cost}
J(\mu,\delta) = \mathbb{E} \left[ \sum_{i=1}^n (X_i-\hat{X}_i)^2\right]
\end{equation}
is minimized.
\end{problem}

\begin{remark} 
Notice that here we are assuming that we know $f_X$, which is not a realistic assumption in many practical scenarios. However, we will first derive our results for this case and lift this assumption later to obtain a completely data-driven design approach.
\end{remark}

\section{Unicast network}\label{sec:unicast}

In this setting, the wireless network behaves as independent links between sensors and their corresponding receivers. However, due to bandwidth constraints, only only link may be active at a time. The scheduler then selects which of the $n$ links to be active, and the remaining links are idle but the observation of a silent symbol still conveys information about the non-transmitted measurements. 

\begin{definition}[Estimation policies for estimation over unicast networks]\label{def:estimator_unicast}
An estimation policy for the $i$-th estimator in the unicast network case is a function parameterized by $\hat{x}_i \in \mathbb{R}$ such that:
\begin{equation}
\delta_i(Y_i) = \begin{cases}
X_i & \text{if} \ \ Y_i=(i,X_i) \\
\hat{x}_i & \text{if} \ \ Y=\varnothing.
\end{cases}
\end{equation}
\end{definition}
Therefore, the collection of estimation policies $\delta$ for \cref{prob:main} is completely characterized by a vector $\hat{x}\in \mathbb{R}^n$, where:
\begin{equation}
\hat{x}\Equaldef (\hat{x}_1,\cdots,\hat{x}_n).
\end{equation}

\begin{theorem}[Difference-of-convex decomposition -- unicast case]\label{thm:DoC_unicast}
If the estimators in \cref{prob:main} use policies of the form in \cref{def:estimator_unicast}, the objective function in \cref{eq:cost} admits the following decomposition as a difference of two convex functions:
\begin{equation}\label{eq:cost_unicast_DoC}
J(\mu^\star_{\delta},\delta) = \mathbb{E}\left[ \sum_{i=1}^n(X_i-\hat{x}_i)^2\right] \\
- \mathbb{E}\left[ \max_{j\in\{1,\cdots,n\}}\Big\{(X_j-\hat{x}_j)^2\Big\}\right],
\end{equation}
where $\mu_{\delta}^\star$ is the optimal scheduler for a fixed collection of estimation policies $\delta$, which is parameterized by the vector $\hat{x}\in \mathbb{R}^n$.
\end{theorem}

\begin{IEEEproof}
Using the estimators in \cref{def:estimator_unicast} and the law of total expectation, the cost function in \cref{eq:cost} can be expressed in integral form as follows:
\begin{equation}
J(\mu,\delta) = \sum_{j=1}^n \int_{\mathbb{R}^n} \Bigg[ \sum_{i\neq j}(x_i-\hat{x}_i)^2\Bigg]\mathbb{I}\big(\mu(x)=j\big)f_X(x)dx.
\end{equation}
For a fixed $\delta$, in other words, for a fixed $\hat{x}\in \mathbb{R}^n$, the optimal scheduler $\mu^\star_{\delta}$ is determined by the following set of inequalities:
\begin{equation}
\mu^{\star}_\delta(x)=j \Leftrightarrow |x_j-\hat{x}_j| \geq |x_\ell-\hat{x}_\ell|, \ \ \ell \in \{1, \cdots, n\}.
\end{equation}
This scheduler leads to the following objective function as a function of $\delta$:
\begin{equation}\label{eq:cost_unicast}
J(\mu^{\star}_{\delta},\delta) = \mathbb{E}\left[ \min_{j\in\{1,\cdots,n\}}\Big\{\sum_{i\neq j}(X_i-\hat{x}_i)^2\Big\}\right].
\end{equation}
The objective function in \cref{eq:cost_unicast} is non-convex due to the $\min\{\cdot\}$ in the argument of the expectation operator. However, the identities hold
\begin{align}
\min_{j}\bigg\{\sum_{i\neq j}(X_i-\hat{x}_i)^2 \bigg\} \nonumber  & =\min_{j} \bigg\{\sum_{i = 1}^n(X_i-\hat{x}_i)^2 - (X_j-\hat{x}_j)^2 \bigg\} \\
 &=  \sum_{i = 1}^n(X_i-\hat{x}_i)^2 + \min_{j} \big\{- (X_j-\hat{x}_j)^2 \big\} \\
 &=  \sum_{i = 1}^n(X_i-\hat{x}_i)^2 - \max_{j} \big\{(X_j-\hat{x}_j)^2 \big\}.
\end{align}
The result follows from the linearity of the expectation operator.
\end{IEEEproof}

The fact that the optimization problem admits a DoC decomposition is attractive because it allows the use of global optimization techniques such as branch-and-bound methods, which are guaranteed to converge to a globally optimal solution \cite{scholz:2012}. However, the convergence of such algorithms is typically very slow for large dimensional optimization problems, which in our case, would be prohibitive in case of a large number of sensors. Therefore, these global optimization techniques would not be suitable for IoT applications. On the other hand, the DoC decomposition also allows for the use of a technique known as the convex-concave procedure (CCP) \cite{Yuille:2003,Shen:2016,Lipp:2016}, which is guaranteed to converge to a locally optimal solution \cite{Sriperumbudur:2009} and often admits simpler and faster implementation. 

\subsection{Convex-concave procedure}

The convex-concave procedure is an optimization technique used to find local minima of non-convex cost functions that admit a DoC decomposition. The advantage of using CCP over a subgradient method is that the CCP makes use of the structure of the cost function, which in certain special cases can lead to very efficient algorithms. Herein, we will apply the CCP to the cost in \cref{eq:cost_unicast_DoC}.

\begin{theorem}
Consider the unconstrained non-convex optimization problem:
\begin{equation}\label{eq:ccp1}
\min_{\hat{x}\in\mathbb{R}^n} J(\hat{x})= F(\hat{x}) - G(\hat{x}), 
\end{equation} 
where
\begin{equation}\label{eq:ccp2}
F(\hat{x}) \Equaldef \mathbb{E}\left[ \sum_{i=1}^n(X_i-\hat{x}_i)^2\right]
\end{equation}
and
\begin{equation}\label{eq:ccp3}
G(\hat{x}) \Equaldef \mathbb{E}\left[ \max_{j\in\{1,\cdots,n\}}\Big\{(X_j-\hat{x}_j)^2\Big\}\right].
\end{equation}

Let $g$ be any subgradient of the function $G$. The dynamical system described by the recursion:
\begin{equation}\label{eq:CCP_unicast}
\hat{x}^{(k+1)} = \frac{1}{2}g(\hat{x}^{(k)})+\mathbb{E}[X]
\end{equation}
converges to a local minimum of $J(\hat{x})$.
\end{theorem}

\begin{IEEEproof}
We will apply the CCP to the optimization problem in \cref{eq:ccp1,eq:ccp2,eq:ccp3}. The CCP consists of approximating the non-convex part of $J$, \emph{i.e.}, $G$, by its affine approximation at a given point $\hat{x}^{(k)} \in \mathbb{R}^n$:
\begin{equation}
G_{\text{affine}}(\hat{x};\hat{x}^{(k)}) \Equaldef G(\hat{x}^{(k)}) + g(\hat{x}^{(k)})^{\mathsf{T}}(\hat{x}-\hat{x}^{(k)}),
\end{equation}
where $g(\hat{x}^{(k)})$ is any subgradient\footnote{A vector $g \in \mathbb{R}^n$ is a subgradient of $f:\mathbb{R}^n\rightarrow \mathbb{R}$ at $x\in \mathbf{dom}\ f$ if for all $z \in \mathbf{dom}\ f$, $$f(z) \geq f(x) + g^{\mathsf{T}}(z-x).$$ } of the function $G$ at the point $\hat{x}^{(k)}$. Then a new point $\hat{x}^{(k+1)}$ is generated according as a solution of a convex optimization problem as follows:
\begin{equation}\label{eq:CCP_general}
\hat{x}^{(k+1)} = \arg \min_{\hat{x}\in \mathbb{R}^n} \Big\{ F(\hat{x}) - G_{\text{affine}}(\hat{x};\hat{x}^{(k)}) \Big\}.
\end{equation}
The unconstrained convex optimization problem in \cref{eq:CCP_general} can be solved by using the first order optimality condition:
\begin{equation}
\nabla ( F(\hat{x}) -G_{\text{affine}}(\hat{x})) \Big|_{\hat{x}=\hat{x}^\star} = 0,
\end{equation}
which in this case has a unique solution. Computing the gradient above at $\hat{x}^{\star}$ yields:
\begin{equation}
2(\hat{x}^{\star}-\mathbb{E}[X]) - g(\hat{x}^{(k)}) = 0.
\end{equation}
Finally, by solving for $\hat{x}^{\star}$, we obtain the following dynamical system:
\begin{equation}
\hat{x}^{(k+1)} = \frac{1}{2}g(\hat{x}^{(k)})+\mathbb{E}[X].
\end{equation}

The sequence of the points generated according to the dynamical system above is guaranteed to converge to one of the local optimizers of $J$. The proof of this fact can be found in \cite{Sriperumbudur:2009}.
\end{IEEEproof}

\subsection{Relationship with subgradient methods}

The dynamical system in \cref{eq:CCP_unicast} is related to subgradient methods of the form:
\begin{equation}\label{eq:subgradient_method}
\hat{x}^{(k+1)} = \hat{x}^{(k)} - \alpha_k j(\hat{x}^{(k)})
\end{equation}
where $j(\hat{x}^{(k)})$ is a subgradient of $J$ at $\hat{x}^{(k)}$. Notice that convergence results for such algorithms exist under the condition that $J$ is a convex function and the step sequence satisfies certain summability conditions\footnote{For example, if the step size sequence $\{\alpha_k\}$ satisfies: $$\sum_{k=0}^{\infty}\alpha_k^2 <\infty \ \ \text{and} \ \ \sum_{k=0}^{\infty}\alpha_k = \infty.$$} that typically imply a very slow convergence rate to a global minimum. There are no guarantees in general that a subgradient method like the one in \cref{eq:subgradient_method} will converge to a local minimum if the the objective function is non-convex. 

The dynamical system from the CCP in \cref{eq:CCP_unicast} is equivalent to:
\begin{equation}
\hat{x}^{(k+1)} = \hat{x}^{(k)} -\frac{1}{2}j(x^{(k)}),
\end{equation}
where 
\begin{equation}
j(x^{(k)}) \Equaldef \nabla F(\hat{x}^{(k)}) - g(\hat{x}^{(k)}).
\end{equation}

The constant step size $\alpha = 0.5$ is highly desirable because it yields a fast convergence to a local minimum despite the fact that the objective function is non-convex. Furthermore, even for convex objectives, the constant step size only guarantees convergence to a point within a fixed gap of the optimal solution \cite{Boyd:2014}.

\subsection{Computing a subgradient}

The dynamical system in \cref{eq:CCP_unicast} relies on the fact that at every time step $k$, we are able to evaluate a subgradient $g$ of the function $G$ defined in \cref{eq:ccp3}. The fact that only a subgradient is required is important because the function $\max$ inside the expectation $G$ is non-differentiable, which may lead to a non-differentiable $G$ depending on the joint PDF $f_X$. Next, we will use weak subgradient calculus to compute a subgradient $g$.

For a fixed vector $x\in \mathbb{R}^n$, define:
\begin{equation}
G(\hat{x};x)\Equaldef \max_{j\in\{1,\cdots,n\}}\Big\{(x_j-\hat{x}_j)^2\Big\},
\end{equation}
and
\begin{equation}
G_j(\hat{x};x) \Equaldef (x_j-\hat{x}_j)^2, \ \ j\in\{ 1,\cdots, n\}.
\end{equation}
Therefore, 
\begin{equation}
G(\hat{x};x) = \max_{j\in\{1,\cdots,n\}}G_j(\hat{x};x)
\end{equation}

The gradient of each $G_j(\hat{x};x)$ is given by
\begin{equation}
\nabla G_j(\hat{x};x) = 
-2(x_j-\hat{x}_j)\mathbf{e}_j,
\end{equation}
where $\mathbf{e}_j$ is the $j$-th canonical basis vector in $\mathbb{R}^n$.


The computation of a subgradient for $G(\hat{x};x)$ is done via an algorithmic procedure, which implements simple a linear search. For a fixed pair of arguments $(\hat{x};x)$, the subgradient is computed as follows:
\begin{equation}
g(x;\hat{x}) = \texttt{subgrad}(x;\hat{x}),
\end{equation}
where \texttt{subgrad} is given in the algorithmic procedure below:

\begin{algorithm}
\caption{}\label{algo:subgrad} 
\begin{algorithmic}[1] 
\Procedure{subgrad}{$\hat{x},x$} \Comment{A subgradient of $G(\hat{x},x)$}

\State $G^{\star} \gets -\infty$

\State $j^{\star} \gets 0$

\For{ $j \in \{1,\cdots,n$\}}\Comment{Linear search}
	
	\State $G \gets G_j(\hat{x};x)$
  
			\If{ $G\geq G^{\star}$ }

				\State $G^{\star} \gets G$

				\State $j^{\star} \gets j$

			\EndIf

\EndFor

\State $g \gets \nabla G_{j^{\star}}(\hat{x};x)$

   \State \textbf{return} $g$\Comment{The vector $g$ is a subgradient of $G$ at $(\hat{x};x)$ }
\EndProcedure
\end{algorithmic}
\end{algorithm}

Finally, weak subgradient calculus states that
\begin{equation}\label{eq:subgrad}
g(\hat{x}) \Equaldef \mathbb{E}\big[g(\hat{x};X) \big],
\end{equation}
belongs to the sub-differential $\partial G(\hat{x})$, where the expectation is taken with respect to the random vector $X$. Thus, \cref{eq:subgrad} is a subgradient of $G$ at $\hat{x}$ \cite{Boyd:2018}.

\begin{remark}
The computational procedure derived from the CCP is simple, but still requires the computation of an $n$-dimensional integral due to the expectation operator. Two things may occur: 1. we know the PDF of the measurement vector $X$, and the dimension $n$ is small enough to allow for efficient numerical computation of the expectation; 2. we do not have access to the PDF or the dimension $n$ is prohibitively large, but we have access to a (sufficiently large) data set. The latter scenario will be explored in \cref{sec:data_driven}.
\end{remark}

\subsection{An illustrative example}

Here we provide an example for the observation-driven scheduling over a unicast network with $n=2$ sensor-estimator pairs. Each sensor observes a component of a bivariate source $X=(X_1,X_2)$. Let $X$ be distributed according to the following mixture of bivariate Gaussians:
\begin{equation}\label{eq:mixture}
X \sim \frac{3}{4} \mathcal{N}\left(\begin{bmatrix} 0 \\ 0 \end{bmatrix} , \begin{bmatrix} 1 & 0 \\ 0 & 1\end{bmatrix} \right) + \frac{1}{4}\mathcal{N}\left(\begin{bmatrix} 4 \\ 2 \end{bmatrix} , \begin{bmatrix} 1 & 0.4 \\ 0.4 & 1\end{bmatrix} \right).
\end{equation}

\begin{figure}[t!]
\centering
\includegraphics[scale=0.45]{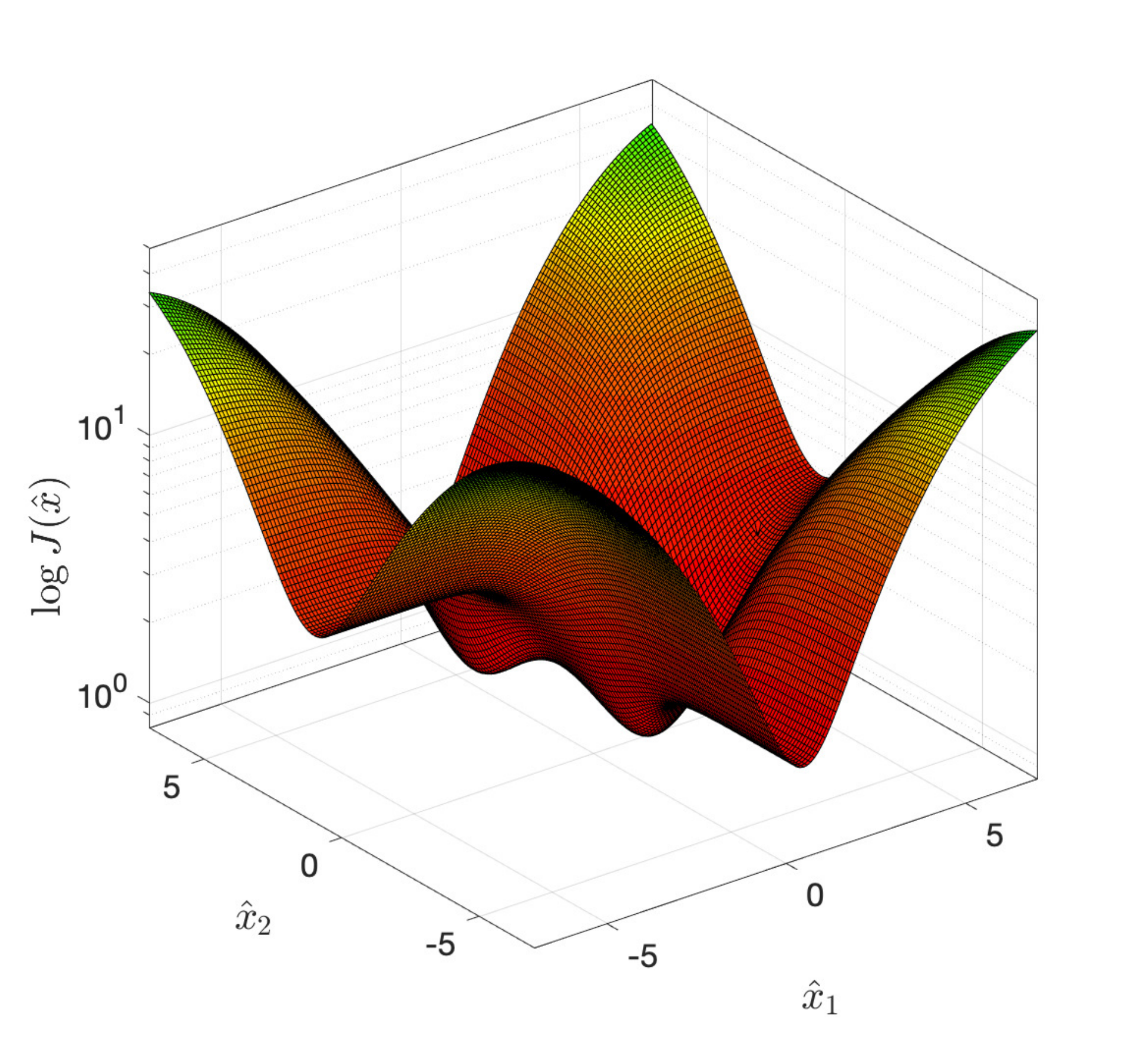}
\caption{Objective function $J(\hat{x})$ in \cref{eq:cost_unicast} for the observations $X_1$ and $X_2$ jointly distributed according to the Gaussian mixture in \cref{eq:mixture}. We have used $\log$ scale to emphasize the two local minima.}
\label{fig:cost_mixture_unicast}
\end{figure}

\begin{figure}
\centering
\includegraphics[width=0.45\textwidth]{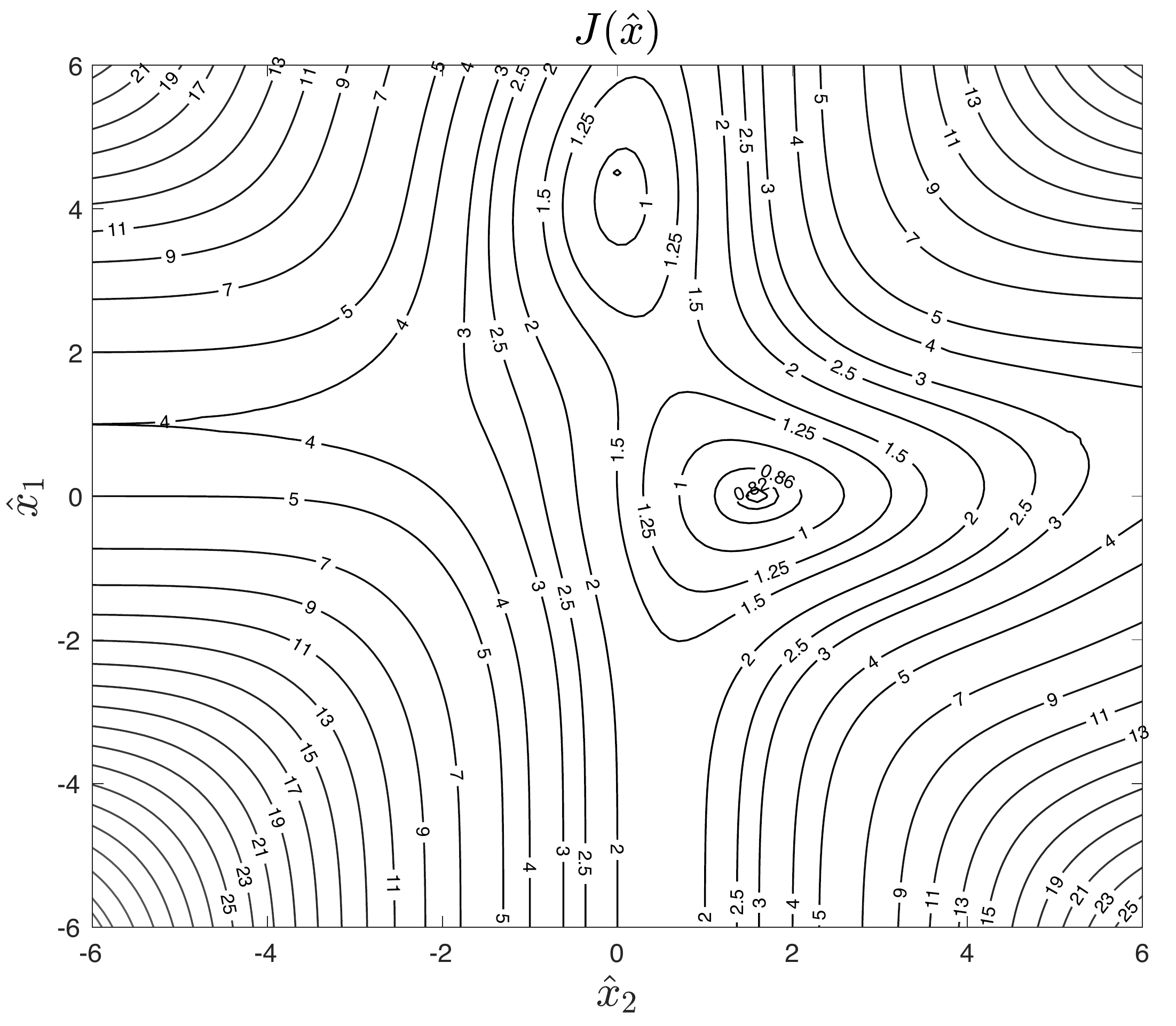}
\caption{Level curves for the objective function $J(\hat{x})$ shown in \cref{fig:cost_mixture_unicast}.}
\label{fig:level_curves_cost_mixture_unicast}
\end{figure}

\Cref{fig:cost_mixture_unicast} shows the objective function in \cref{eq:cost_unicast} for the PDF in \cref{eq:mixture} (shown in logarithmic scale to emphasize its two local minima) and its level curves are shown in \cref{fig:level_curves_cost_mixture_unicast}. As it was expected, this objective function is non-convex.

Assuming that we did not know the number of local minima, we used the algorithm in \cref{eq:CCP_unicast} for $1000$ random initial conditions $\hat{x}^{(0)}\in\mathbb{R}^2$, and kept the resulting $\hat{x}^{\star}$ with the best value. In our case, we obtained: 
\begin{equation}
\hat{x}^{\star} = \begin{bmatrix}
 0.0045 \\ 1.5900
\end{bmatrix}
\end{equation}
with an associated value of
\begin{equation}
J(\hat{x}^\star) = 0.8065.
\end{equation}
Therefore, the optimal scheduler is given by 
\begin{equation}
\mu^{\star}(x)= \begin{cases}1 & \ \ \text{if} \ \ |x_1- 0.0045| \geq |x_2-1.5900| \\
2 & \ \ \text{otherwise},
\end{cases}
\end{equation}
the optimal estimators are given by
\begin{equation}
\delta^{\star}_1(Y_1) = \begin{cases}
X_1 & \text{if} \ \ Y_1=(1,X_1) \\
0.0045 & \text{if} \ \ Y_1=\varnothing
\end{cases}
\end{equation}
and
\begin{equation}
\delta^{\star}_2(Y_2) = \begin{cases}
X_2 & \text{if} \ \ Y_2=(2,X_2) \\
1.5900 & \text{if} \ \ Y_1=\varnothing.
\end{cases}
\end{equation}

In order to compare the performance of this observation-driven scheduler, consider a blind-scheduler, $\mu^{\textrm{blind}}$, which gives channel access to the sensor with the largest variance. The corresponding blind-estimators $\delta^{\textrm{blind}}$ output the expected value of the unobserved random variable, i.e.,
\begin{equation}
\mu^{\textrm{blind}}(x) = \arg \max_{i\in\{1,2\}}\mathrm{Var}(X_i)
\end{equation} 
and
\begin{equation}
\delta^{\textrm{blind}}_i(Y_i) = \begin{cases}
X_i & \text{if} \ \ Y_i=(i,X_i) \\
\mathbb{E}[X_i] & \text{if} \ \ Y_i=\varnothing.
\end{cases}
\end{equation} 

In this example, the performance of the blind scheduler is:
\begin{equation}
J(\mu^{\textrm{blind}},\delta^{\textrm{blind}}) = \min\{4,1.75\} = 1.75.
\end{equation} 
Notice that the performance of the observation-driven scheduler in this case is approximately $54\%$ better than the blind-scheduler.


\section{Broadcast network}\label{sec:broadcast}

When the wireless network is of the broadcast type, all the estimators, receive the same signal. This signal is then used as side information to estimate the non-received random variables. Given that $U=j$, the received signals at the estimators are: 
\begin{equation}
Y_i = (j,X_j), \ \ i\in\{1,\cdots,n\}.
\end{equation}
In this case, $X_j$ serves as side information for the estimates $\hat{X}_i$, $i\neq j$. This must be the case even if the sensors make mutually independent observations.

\begin{proposition}
Consider \cref{prob:main} over a broadcast network. Let $i,j \in \{1,\cdots,n\}$ such that $i\neq j$. For a fixed scheduling policy $\mu$, the optimal estimator $\delta_{\mu,i}^{\star}$ is of the following form:
\begin{equation}
\delta_{\mu,i}^{\star}(Y_i) = \begin{cases}
X_i & \ \ Y_i = (i,X_i) \\
\eta_{ij}(X_j) & \ \  Y_j = (j,X_j),
\end{cases}
\end{equation}
where $\eta_{ij}$ are functions that depend implicitly on $\mu$.  
\end{proposition}

\begin{IEEEproof}
For a fixed scheduling policy $\mu$, the mean-squared error objective function implies that the optimal estimator is the conditional mean of the measurement given the channel output, i.e., for $U=j$,
\begin{equation}
\delta_{i,\mu}^{\star}(j,x_j) = \mathbb{E}\big[X_i \mid \mu(X)=j,X_j=x_j \big]. 
\end{equation}
If $i=j$, then
\begin{equation}
\mathbb{E}\big[X_i \mid \mu(X)=i,X_i=x_i \big] = x_i.
\end{equation}
If $i\neq j$, then
\begin{equation}
\mathbb{E}\big[X_i \mid \mu(X)=j,X_j=x_j \big] \Equaldef \eta_{ij}(x_j).
\end{equation}
\end{IEEEproof}

\begin{remark}
Without making any assumptions on the probabilistic model or the scheduler there is nothing we can say about the structure of the optimal representation functions $\eta_{ij}$. In fact, even if the observations are jointly Gaussian, the optimal representation functions can be nonlinear \cite{Vasconcelos:2019b}. 
\end{remark}

In order to obtain a tractable finite dimensional optimization problem we will constrain the estimators for a broadcast network to the affine class.

\begin{definition}[Policies for estimation over boradcast networks]\label{def:estimator_broadcast}
An estimation policy for the $i$-th estimator in the broadcast network case is a function parameterized by weights $w_{ij} \in \mathbb{R}$ and biases $b_{ij}\in\mathbb{R}$, such that:\begin{equation}
\delta_i(Y_i) = \begin{cases}
X_i & \text{if} \ \ Y_i = (i,X_i) \\
w_{ij}X_j + b_{ij} & \text{if} \ \ Y_i = (j,X_j) \ \text{and} \ j\neq i.
\end{cases} 
\end{equation}
\end{definition}


We are trading off optimality for tractability by constraining the class of estimators to be affine, and performing the optimization within that class. Also notice that there is a considerable increase in the complexity of the estimators: The total number of optimization variables in this version of \Cref{prob:main} is equal to the number of parameters used to describe all the estimators. In this case, this number is:
\begin{equation}
d = 2(n-1)n.
\end{equation} 
Therefore, the number of variables scales quadratically with the number of sensors, as opposed to the the linear number of variables in the unicast case. Nevertheless, the number of variables in our algorithm scales polynomially in the number of sensors, which is still manageable for applications with a large number of sensors, such as IoT.

Therefore, the collection of estimation policies $\delta$ for \cref{prob:main} is characterized by $\theta\in\mathbb{R}^d$:
\begin{equation}
\theta \Equaldef \mathrm{vec}(\theta_1,\cdots,\theta_n),
\end{equation}
where
\begin{equation}
\theta_j\Equaldef \mathrm{vec}\Big( \Big\{ \begin{bmatrix}w_{ij} \\ b_{ij}\end{bmatrix}, i\neq j \Big\}\Big).
\end{equation}

\begin{theorem}[Difference-of-convex decomposition -- broadcast case]\label{thm:DoC_broadcast}
If the estimators in \cref{prob:main} use policies of the form in \cref{def:estimator_broadcast}, the objective function in \cref{eq:cost} admits the following decomposition as a difference of two convex functions:
\begin{equation}
J(\mu_{\delta}^{\star},\delta) = \mathbb{E} \Bigg[ \sum_{\ell=1}^n \sum_{i \neq \ell} (X_i - (w_{i\ell}X_\ell+b_{i\ell}) )^2\\ -\max_{j \in \{1,\cdots,n\}} \Big\{ \sum_{(i,\ell) \neq (j,j)} \big(X_i-(w_{i\ell}X_\ell+b_{i\ell})\big)^2\Big\}\Bigg].
\end{equation}
\end{theorem}

\begin{IEEEproof}
For a fixed collection of estimation policies of the form given in \cref{def:estimator_broadcast}, i.e. for a fixed vector $\theta\in \mathbb{R}^d$, and using the law of total expectation, the cost function in \cref{eq:cost} can be expressed in integral form as follows:
\begin{equation}
J(\mu,\delta) = \sum_{j=1}^n \Bigg[ \int_{\mathbb{R}^n} \Big(\sum_{i\neq j}\big( x_i - (w_{ij}x_j-b_{ij}) \big)^2 \Big)\times \\
\mathbb{I}(\mu(x)=j)f_X(x)dx \Bigg].
\end{equation}
The optimal scheduling policy $\mu_\delta^{\star}(x) = j$ if and only if the following set of inequalities are satisfied:
\begin{equation}
\sum_{i\neq j} \big(x_i-(w_{ij}x_j+b_{ij})\big)^2 < \sum_{i\neq \ell} \big(x_i-(w_{i\ell}x_j+b_{i\ell})\big)^2, \ \ \ell\neq j.  
\end{equation}

Using this scheduler, we may rewrite the optimization problem as a function of the parameters of the estimators, $\theta$. Thus,
\begin{equation}
J(\mu_{\delta}^{\star},\delta) = \mathbb{E}\Bigg[ \min_{j\in\{1,\cdots,n\}} \Big\{ \sum_{i\neq j} \big(X_i-(w_{ij}X_j+b_{ij})\big)^2\Big\}\Bigg].
\end{equation}
The following identity holds:
\begin{equation}
\sum_{i\neq j} \big(X_i-(w_{ij}X_j-b_{ij})\big)^2 = \sum_{\ell=1}^n \sum_{i\neq \ell} \big(X_i-(w_{i\ell}X_\ell+b_{i\ell})\big)^2 \\ -  \sum_{(i,\ell)\neq (j,j)}\big(X_i-(w_{i\ell}X_\ell+b_{i\ell})\big)^2.
\end{equation}
Finally,
\begin{equation}
\min_j\sum_{i\neq j} \big(X_i-(w_{ij}X_j+b_{ij})\big)^2 = \sum_{\ell=1}^n \sum_{i\neq \ell} \big(X_i-(w_{i\ell}X_\ell+b_{i\ell})\big)^2 \\ - \max_j  \sum_{(i,\ell)\neq (j,j)}\big(X_i-(w_{i\ell}X_\ell+b_{i\ell})\big)^2.
\end{equation}
\end{IEEEproof}
\begin{remark}
Notice that the DoC decomposition in the broadcast case is not as neat as in the unicast case. The reason is that for each $X_j$, the estimator uses a different set of parameters $w_{ij},b_{ij}$. However, the decomposition in \cref{thm:DoC_broadcast} is just as useful as the one in \cref{thm:DoC_unicast}.
\end{remark}

\subsection{Convex-concave procedure}


For the remainder of this section, we will assume that $n=2$. The equations for $n>2$ are presented in the appendix.

The parameter vector $\theta$ which specifies the affine estimators $\delta_1$ and $\delta_2$ is:
\begin{equation}
\theta = (w_{21},b_{21},w_{12},b_{12}).
\end{equation}

\begin{theorem}
Consider the unconstrained non-convex optimization problem:
\begin{equation}\label{eq:ccp_broad_1}
\min_{\theta\in\mathbb{R}^4} J(\theta)= F(\theta) - G(\theta), 
\end{equation} 
where
\begin{equation}\label{eq:ccp_broad_2}
F(\theta) \Equaldef \mathbb{E} \Big[  (X_1 - (w_{12}X_2+b_{12}) )^2 + \\ (X_2 - (w_{21}X_1+b_{21}) )^2 \Big]
\end{equation}
and
\begin{equation}\label{eq:ccp_broad_3}
G(\theta) \Equaldef \mathbb{E} \Big[  \max \Big\{ \big(X_1-(w_{12}X_2+b_{12})\big)^2,\\\big(X_2-(w_{21}X_1+b_{21})\big)^2\Big\} \Big].
\end{equation}

Let $g$ be any subgradient of the function $G$. One such subgradient is given by 
\begin{equation}\label{eq:subgradient_broadcast}
g(\theta) = -2\mathbf{E}\begin{bmatrix}

X_1(X_2-w_{21}X_1-b_{21})\mathbb{I}(|X_1-w_{12}X_2-b_{12}|<|X_2-w_{21}X_1-b_{21}|)\\
\ \ \ \  (X_2-w_{21}X_1-b_{21})\mathbb{I}(|X_1-w_{12}X_2-b_{12}|<|X_2-w_{21}X_1-b_{21}|)\\
X_2(X_1-w_{12}X_2-b_{12})\mathbb{I}(|X_1-w_{12}X_2-b_{12}|\geq|X_2-w_{21}X_1-b_{21}|)\\
\ \ \ \ (X_1-w_{12}X_2-b_{12})\mathbb{I}(|X_1-w_{12}X_2-b_{12}|\geq|X_2-w_{21}X_1-b_{21}|)
\end{bmatrix}.
\end{equation}

Let $\mathbf{A}$ and $\mathbf{b}$ be defined as:  
\begin{equation}\label{eq:matrix}
\mathbf{A} \Equaldef 2\begin{bmatrix}
\mathbb{E}[X_1^2] & \mathbb{E}[X_1] & 0 & 0 \\
\mathbb{E}[X_1] & 1 & 0 & 0 \\
0 & 0 & \mathbb{E}[X_2^2] & \mathbb{E}[X_2]\\
0 & 0 & \mathbb{E}[X_2] & 1
\end{bmatrix}
\end{equation}
and
\begin{equation}
\mathbf{b} \Equaldef 2 \begin{bmatrix}
\mathbb{E}[X_1X_2] \\
\mathbb{E}[X_2] \\
\mathbb{E}[X_1X_2] \\
\mathbb{E}[X_1]
\end{bmatrix}.
\end{equation}

The dynamical system described by the recursion:
\begin{equation}\label{eq:CCP_broadcast}
\theta^{(k+1)} = \mathbf{A}^{-1}\big(g(\theta^{(k)})+ \mathbf{b}\big)
\end{equation}
converges to a local minimum of $J(\theta)$.
\end{theorem}

\begin{remark}
Under the assumption that the observations at the sensors $X_1$ and $X_2$ are random variables with finite first and second moments, the matrix $\mathbf{A}$ is always invertible.
\end{remark}

\begin{IEEEproof}
Using the CCP to the minimization problem in \cref{eq:ccp_broad_1,eq:ccp_broad_2,eq:ccp_broad_3}, we have
\begin{equation}\label{eq:CCP_general_broadcast}
\hat{\theta}^{(k+1)} = \arg \min_{\theta\in \mathbb{R}^4} \Big\{ F(\theta) - G_{\text{affine}}(\theta;\theta^{(k)}) \Big\},
\end{equation}
where
\begin{equation}
G_{\text{affine}}(\theta;\theta^{(k)}) \Equaldef G(\theta^{(k)}) + g(\theta^{(k)})^{\mathsf{T}}(\theta-\theta^{(k)}).
\end{equation}
The unconstrained convex optimization problem in \cref{eq:CCP_general_broadcast} can be solved by using the first order optimality condition:
\begin{equation}
\nabla \big( F(\theta) -G_{\text{affine}}(\theta;\theta^{(k)})\big) \Big|_{\theta=\theta^\star} = 0,
\end{equation}
which in this case has a unique solution. Computing the gradient above at $\hat{x}^{\star}$ yields:
\begin{equation}
\mathbf{A}\theta^{\star}-\mathbf{b} - g(\theta^{(k)}) = 0.
\end{equation}
Solving for $\theta^{\star}$ gives the dynamical system in \cref{eq:CCP_broadcast}. The convergence to a local minimum is guaranteed by the CCP.
\end{IEEEproof}









\begin{remark}
The computational bottleneck in our algorithm comes from the fact that in order to compute the subgradient in \cref{eq:subgradient_broadcast} we need to compute 2-dimensional integrals of arguments that involve indicator functions. These are numerically hard to deal with and may lead to slow convergence rates, and oftentimes, the integral may not converge at all, leading to poor performance. The situation is further complicated when the number of sensor-estimator pairs is large. However, the most important observation is that the overall structure of the algorithm does not depend on the distribution of the data.
\end{remark}

\subsection{Relationship with subgradient methods}

The algorithm of \cref{eq:subgradient_broadcast} can also be put in a form that resembles a subgradient method as follows:
\begin{equation}
\theta^{(k+1)} = \theta^{(k)} - \mathbf{A}^{-1}j(\theta^{(k)}).
\end{equation}

As opposed to the algorithm obtained for unicast networks, there is not a scalar step size. The subgradient $j(\theta^{(k)})$ is instead multiplied by the matrix $\mathbf{A}^{-1}$. Therefore, the step size corresponds to spectral radius of $\mathbf{A}^{-1}$, and it is still constant.

\begin{corollary}
The step size $\alpha$ of the algorithm in \cref{eq:subgradient_broadcast} is the sepctral radius of the inverse of $\mathbf{A}$ defined in \cref{eq:matrix}:
\begin{equation}
\alpha  \Equaldef 
\rho(\mathbf{A}^{-1}).
\end{equation}
\end{corollary} 

\subsection{An illustrative example}

Consider the observation driven scheduling over a broadcast network with $n=2$ sensor-estimator pairs. Each sensor observes a component of a bivariate source $X=(X_1,X_2)$. Let $X$ be distributed according to the same mixture of bivariate Gaussians of \cref{eq:mixture}. Running the algorithm in \cref{eq:CCP_broadcast} for $1000$ different initial conditions, $\theta^{(0)}$, and keeping the solution with the best value, we obtain the following 
\begin{equation}
\theta^\star =  (0.4238, \ 0.2151 ,\ -0.2390, \ 0.0624 )
\end{equation}
with
\begin{equation}
J(\theta^\star) = 0.5276.
\end{equation}
Therefore, the optimal scheduler is given by 
\begin{equation}
\mu^{\star}(x)= \begin{cases}1 & \ \ \text{if} \ \ |x_1 + 0.2390x_2 - 0.0624| \geq |x_2-0.4238x_1 - 0.2151|  \\
2 & \ \ \text{otherwise},
\end{cases}
\end{equation}
and the optimal estimators are given by
\begin{equation}
\delta^{\star}_1(Y_1) = \begin{cases}
X_1 & \text{if} \ \ Y_1=(1,X_1) \\
-0.2390X_2 + 0.0624 & \text{if} \ \ Y_1=(2,X_2)
\end{cases}
\end{equation}
and
\begin{equation}
\delta^{\star}_2(Y_2) = \begin{cases}
X_2 & \text{if} \ \ Y_2=(2,X_2) \\
0.4238X_1 + 0.2151 & \text{if} \ \ Y_2=(1,X_1).
\end{cases}
\end{equation}

Comparing the performance of the optimal scheme obtained for a unicast network with the one obtained here for the broadcast network, we observe an improvement of $34.58\%$. This is possible due to the additional side information provided by the broadcast channel to all the estimators at every transmission. However, this comes at the price of a more complex optimization problem involving a larger number of optimization variables. 

\section{Data-driven sensor scheduling}\label{sec:data_driven}

With the machinery developed in \cref{sec:unicast,sec:broadcast}, we are finally ready to address the case in which the PDF is unknown but a dataset $\mathcal{D}$ is available to the designer. The fact that all the theoretical results and associated algorithms hold irrespective of the joint PDF $f_X$ is very important. However, since the PDF is not available to us, we cannot compute expectations. Here we propose a heuristics that consists of replacing the expectations by their corresponding empirical means computed based on $\mathcal{D}$. For the remainder of this section, we will assume the design of schedulers and estimators for broadcast networks with $n=2$. The design for unicast networks would follow the same steps and is omitted for brevity.

\subsection{Learning framework}

Consider a data set of size $N>1$:
\begin{equation}
\mathcal{D} = \big\{x_1(k),\cdots,x_n(k)\big\}_{k=1}^N
\end{equation}
Define the empirical risk as:
\begin{equation}
J_{\mathcal{D}}(\theta) = F_{\mathcal{D}}(\theta) - G_{\mathcal{D}}(\theta),
\end{equation}
where
\begin{equation}
F_{\mathcal{D}}(\theta) \Equaldef \frac{1}{N}\sum_{k=1}^N\Big[  (x_1(k) - w_{12}x_2(k)-b_{12} )^2 + \\ (x_2(k) - w_{21}x_1(k)-b_{21} )^2 \Big]
\end{equation}
and
\begin{equation}
G_{\mathcal{D}}(\theta) \Equaldef \frac{1}{N}\sum_{k=1}^N \Big[  \max \Big\{ \big(x_1(k)-w_{12}x_2(k)-b_{12}\big)^2, \\ \big(x_2(k)-w_{21}x_1(k)-b_{21}\big)^2\Big\} \Big].
\end{equation}

The CCP operates exactly the same as before, but with the advantage that computing a subgradient involves evaluating an empirical mean rather than computing an integral, which can be done far more efficiently and faster than solving an integral, even if we knew the PDF. The (approximate) CCP algorithm becomes:
\begin{equation}\label{eq:CCP_broadcast_approx}
\theta^{(k+1)} = \mathbf{A}_{\mathcal{D}}^{-1}g_{\mathcal{D}}(\theta^{(k)}) + \mathbf{b}_{\mathcal{D}}, 
\end{equation}
where
\begin{equation}
\mathbf{A}_{\mathcal{D}} = \frac{2}{N} \sum_{k=1}^N\begin{bmatrix}
x_1^2(k) & x_1(k) & 0 & 0 \\
x_1(k) & 1 & 0 & 0 \\
0 & 0 & x_2^2(k) & x_2^2(k)\\
0 & 0 & x_2(k) & 1
\end{bmatrix},
\end{equation}
\begin{equation}
\mathbf{b}_{\mathcal{D}} =  \frac{2}{N} \sum_{k=1}^N\begin{bmatrix}
x_1(k)x_2(k) \\
x_2(k) \\
x_1(k)x_2(k) \\
x_1(k)
\end{bmatrix},
\end{equation}
and $g_{\mathcal{D}}$ is any subgradient of $G_\mathcal{D}$.

\begin{figure}[ht!]
\centering
\includegraphics[width=0.45\textwidth]{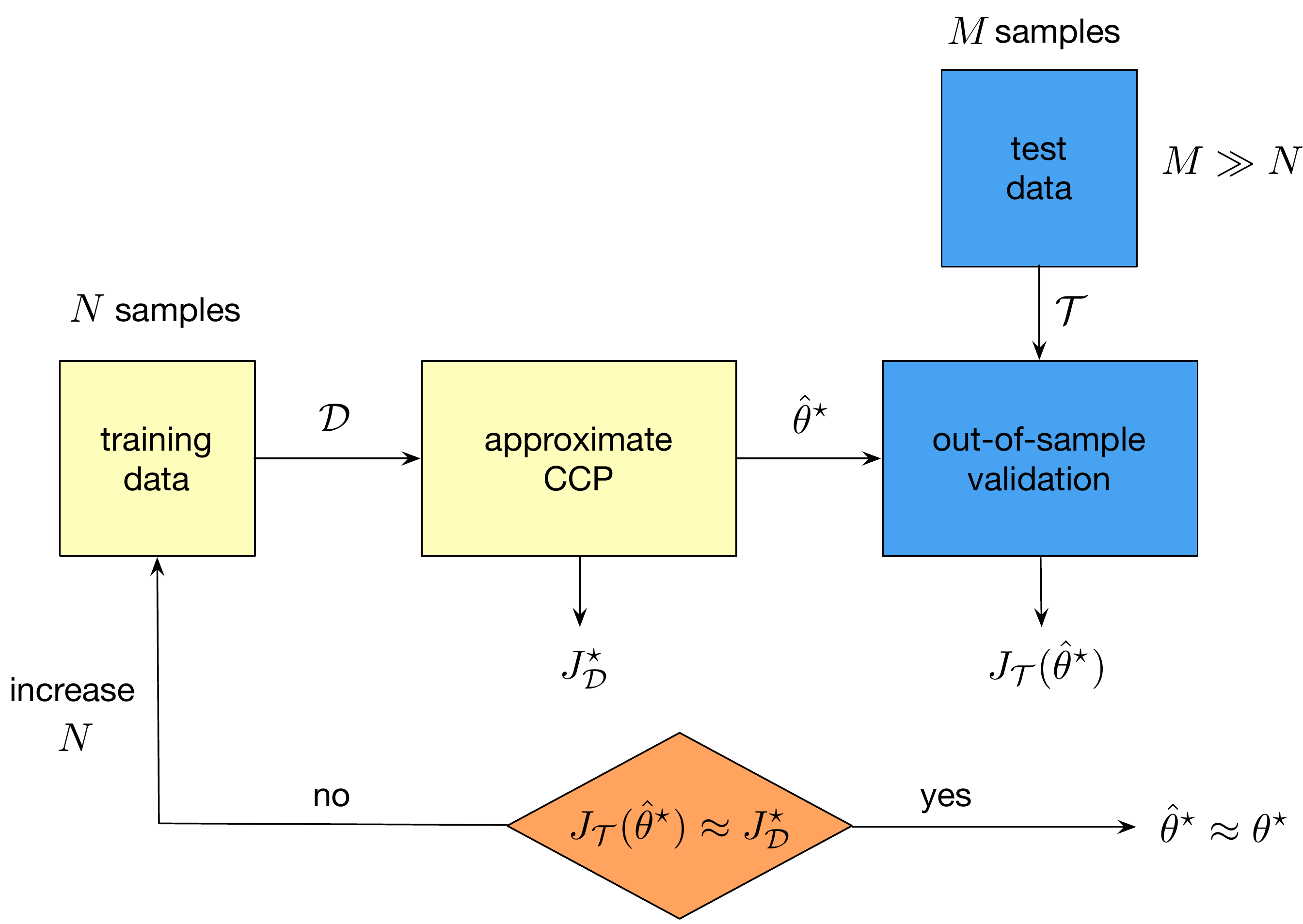}
\caption{Block diagram of the framework used to learn the parameters of the optimal scheduler and associated estimators.}
\label{fig:ML}
\end{figure}

The algorithm above converges to a local minimum $\hat{\theta}^\star$ of the empirical risk $J_{\mathcal{D}}$, and not of the original cost $J$. But, if $N$ is sufficiently large, the empirical risk $J_{\mathcal{D}}$ will be close to $J$ and to the solution obtained through our algorithm $\hat{\theta}^\star$  that would have been obtained if we had access to the unknown PDF, $f_X$. By assumption, we cannot verify exactly how far the solution obtained from the approximate CCP computed using the (training) dataset $\mathcal{D}$, is from the true solution $\theta^\star$. Instead, we perform out-of-sample validation by evaluating the empirical risk using an independent test dataset $\mathcal{T}$ of size $M$, where $M\gg N$. If the value of the empirical risk $J_{\mathcal{T}}(\hat{\theta}^\star)$ is approximately equal to $J_{\mathcal{D}}(\hat{\theta}^\star)$, we declare success and that we have learned the parameters that characterize the optimal scheduler and estimators $\theta^\star$ with some degree of confidence. If $J_{\mathcal{T}}(\hat{\theta}^\star)$ is not approximately equal to $J_{\mathcal{D}}(\hat{\theta}^\star)$, we declare failure and are forced to increase the size of the training data, and repeat the process. This learning framework is based on \cite{Abu-Mostafa:2012} and illustrated in the block diagram in \cref{fig:ML}.

\subsection{An illustrative example}

Consider the a dataset $\mathcal{D}$ consisting of $N=10,000$ samples independent and identically distributed according to the Gaussian mixture model of \cref{eq:mixture}. Using the algorithm in \cref{eq:CCP_broadcast_approx}, we obtained the following solution:
\begin{equation}
\hat{\theta}^\star = (  0.4209,
\    0.2465,\  -0.1905,
\    0.1213
  ).
\end{equation}

There are two values associated with this solution: 
\begin{enumerate}[1.] 

\item The value of the empirical risk (known):
\begin{equation}
J_{\mathcal{D}}(\hat{\theta}^\star) = 0.5257.
\end{equation}

\item  The value of the population risk (unknown):
\begin{equation}
J(\hat{\theta}^\star) = 0.5286.
\end{equation}

\end{enumerate}
If we could compare them directly, we would be able to decide whether this level of performance is acceptable for our application, and adjust the size of the training dataset accordingly to obtain a better approximation, if one is desired. 

Generating an independent test dataset $\mathcal{T}$ consisting of $M=100,000$ samples from the same distribution yields a third value of the objective:
\begin{equation}
J_{\mathcal{T}}(\hat{\theta}^\star) = 0.5250.
\end{equation} 

If there is a single dataset $\mathcal{T}$ to validate the solution, the designer has to make a subjective decision if $J_{\mathcal{T}}(\hat{\theta}^\star)$ is approximately equal to $J_{\mathcal{D}}(\hat{\theta}^\star)$. In this case, the difference is of $0.13\%$, which effectively means that we have learned the parameters of the optimal scheduler and estimators. If we are allowed to run multiple independent experiments to generate test datasets, a more refined analysis using the empirical distribution of $J_\mathcal{T}(\hat{\theta}^\star)$ can be done.

\begin{figure}[b!]
\centering
\includegraphics[width=0.45\textwidth]{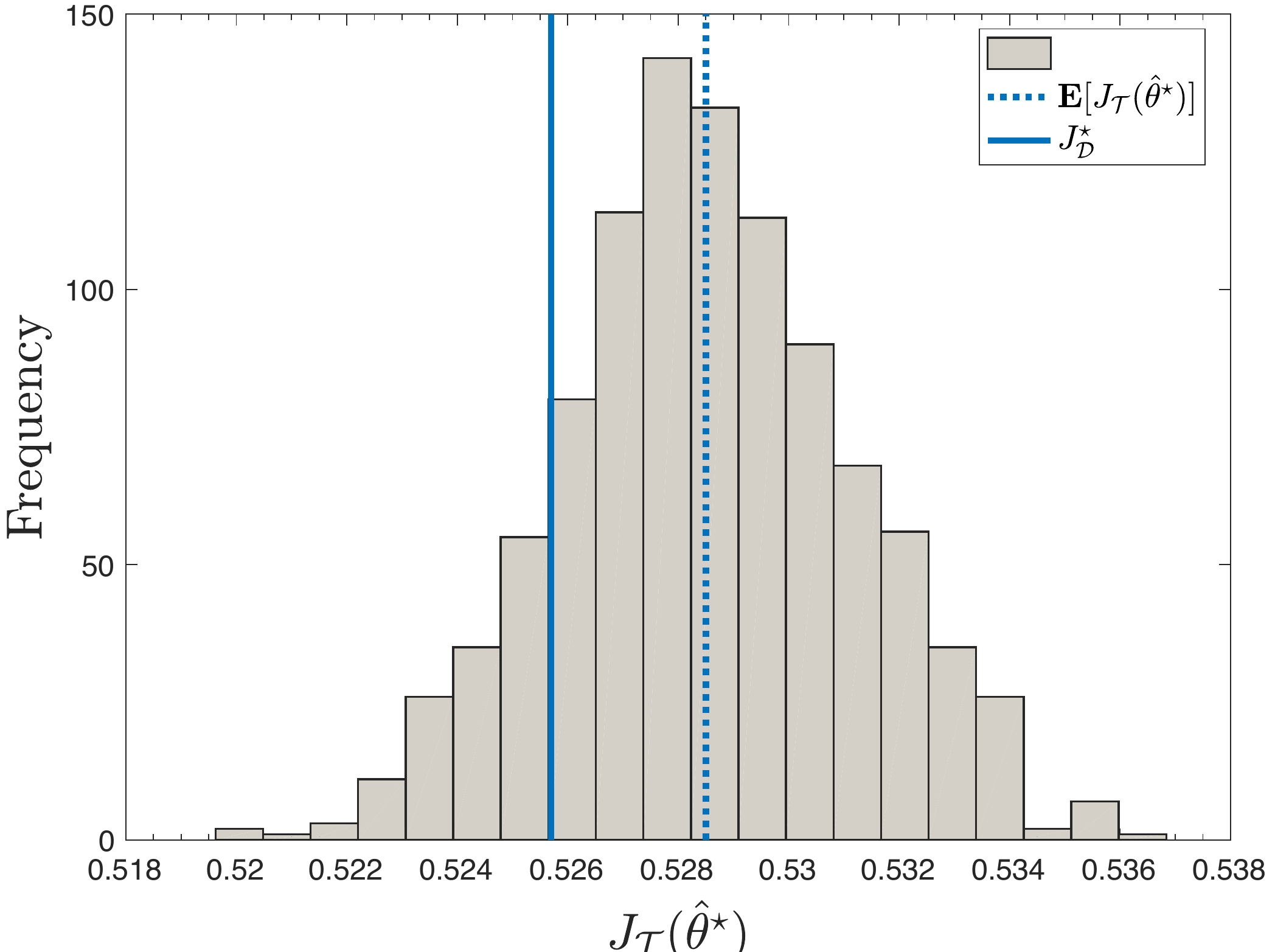}
\caption{Empirical distribution of the random variable $J_{\mathcal{T}}(\hat{\theta}^{\star})$ obtained from $1000$ independent experiments.}
\label{fig:histogram}
\end{figure}

Suppose a number of $1,000$ independent test experiments can be generated, each resulting in a different test dataset $\mathcal{T}$. Computing the value of the objective at $\hat{\theta}^\star$ for each $\mathcal{T}$ yields the histogram shown in \cref{fig:histogram}. First notice that $J_{\mathcal{D}}(\hat{\theta}^\star)$ is within $0.53\%$ from $\mathbb{E}[J_{\mathcal{D}}(\hat{\theta}^\star)]=0.5285$, which is the mean of all the values for each test dataset $\mathcal{T}$, confirming we are indeed at a solution close to the true optimal $\theta^\star$. 

\begin{figure}[ht!]
\centering
\includegraphics[width=0.45\textwidth]{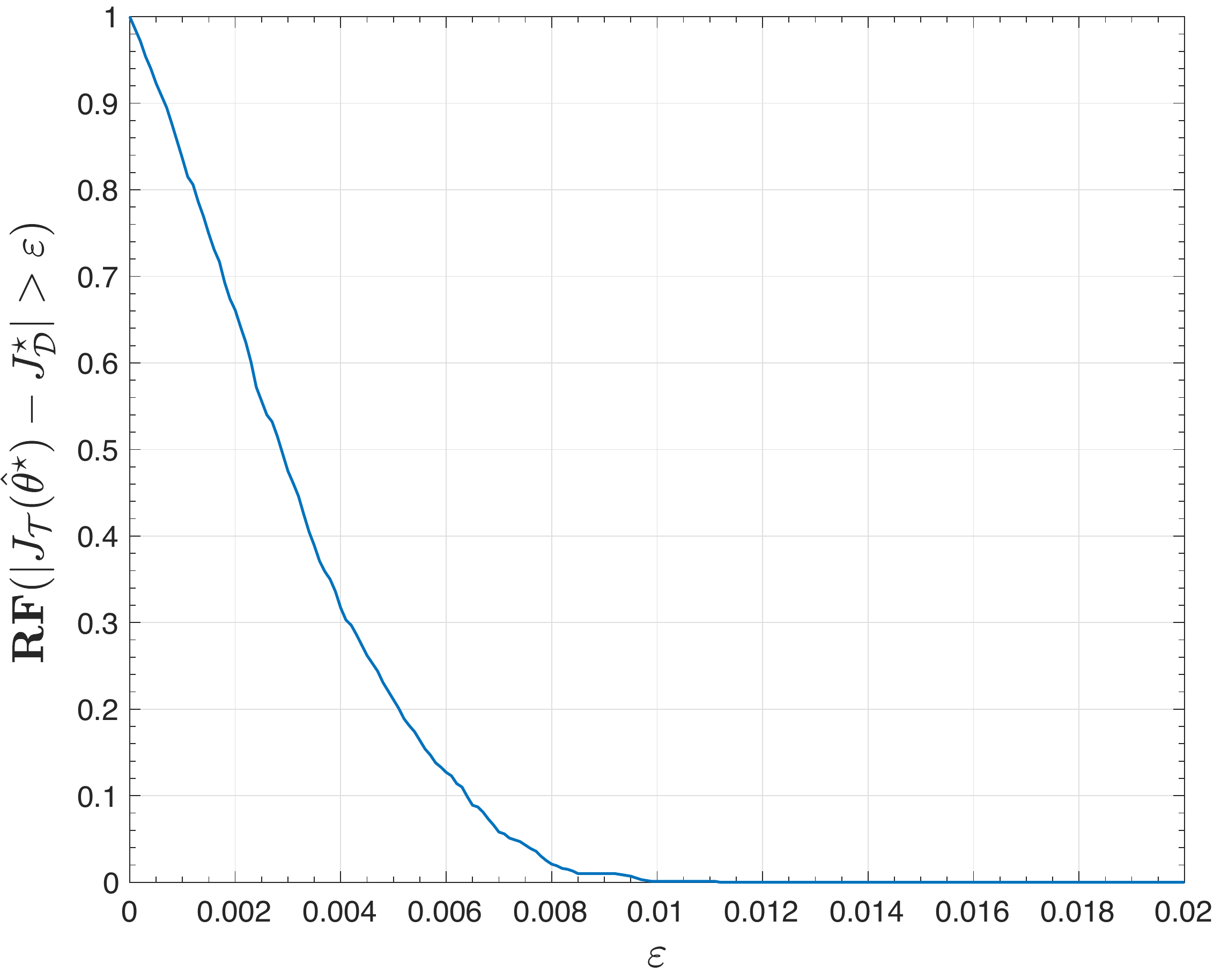}
\caption{Relative frequency of the distance between $J_{\mathcal{T}}(\hat{\theta}^{\star})$ and $J_{\mathcal{D}}(\hat{\theta}^{\star})$ being larger than a constant $\varepsilon$.}
\label{fig:RF}
\end{figure}

Lastly, from the empirical distribution in \cref{fig:histogram}, we may also compute the relative frequency of the distance between $J_{\mathcal{T}}(\hat{\theta}^{\star})$ and $J_{\mathcal{D}}(\hat{\theta}^{\star})$ being greater than a constant $\varepsilon$. \Cref{fig:RF} shows that $J_{\mathcal{T}}(\hat{\theta}^\star)$ is indeed concentrated around $J_{\mathcal{D}}(\hat{\theta}^{\star})$, and we can say that with a very high degree of confidence that $\hat{\theta}^{\star}$ is close to the optimal solution $\theta^\star$.

\section{Conclusions and future work}

This article aims at establishing the foundations for scheduling and estimation of sensor measurements when information about the probabilistic model of the problem is imprecise, missing or incomplete.
We have considered the design of observation-driven schedulers for a remote sensing system for which the random measurements at the sensors are jointly distributed according to an unknown PDF. Such situations occur in many practical applications where the probabilistic model is not known \textit{a priori} or whose underlying physical processes that generate the data are difficult to obtain. We first derive results and accompanying algorithms that hold for an arbitrary joint PDF, and later we use them in a data-driven framework where training and test datasets are available to design the parameters of a scheduler with performance close to the optimal ones.

The framework proposed herein assumes that the wireless network can be of two types: unicast or broadcast. For each case, we show that the optimization problem is non-convex, but admits a useful difference-of-convex decomposition, which allows us to use the convex-concave procedure to obtain very efficient descent algorithms that are guaranteed to converge to a local minimum of the objective function. The structure of both algorithms is independent of the measurements' joint PDF and can be approximated using data by replacing expectations with their corresponding empirical means. Moreover, both algorithms can be mapped as subgradient methods with constant step sizes with guaranteed convergence properties, which are not necessarily convergent if used on non-convex objective functions.

There are many opportunities for future research that branch out from this work. One possible problem is to devise an online learning scheme where the data becomes available one sample at a time to the system designer, which adaptively reconfigures the scheduling and estimation decision rules over time, instead of using batches of data as it was done here. Another line of work is to use concentration inequalities to obtain  performance guarantees as a function of the size of the datasets used for training. It would also be interesting to assume other classes of parametrizable nonlinear estimators for the optimization problem over broadcast networks. For example, we are interested in  the question: can we train neural networks to serve as estimation policies at the estimators? Moreover, can we find neural network architectures that will preserve a difference-of-convex decomposition and take advantage of the convex-concave procedure? Finally, we suggest an entirely new framework where data is used in a distributionally robust framework, where a set of PDFs consistent with the observed data is constructed and a minimax optimization problem is solved as in \cite{Cherukuri:2018}.

\appendices

\section{Vectors and matrices for the broadcast case with $n>2$}

The results in \cref{sec:broadcast} hold for an arbitrary number of sensors. In this appendix we show the structure of the matrices and vectors that define the CCP algorithm in the general case. Recall that:
\begin{equation}
\theta^{(k+1)} = \mathbf{A}^{-1}\big(g(\theta^{(k)})+ \mathbf{b}\big),
\end{equation}
with
\begin{equation}
\mathbf{A} = 2\cdot\mathrm{diag}(\mathbf{A}_1,\cdots,\mathbf{A}_n),
\end{equation}
where
\begin{equation}
\mathbf{A}_j = \begin{bmatrix} \mathbb{E}[X_j^2] & \mathbb{E}[X_j] \\ \mathbb{E}[X_j] & 1 \end{bmatrix};
\end{equation}
and
\begin{equation}
\mathbf{b} = \mathrm{vec}(\mathbf{b}_1,\cdots,\mathbf{b}_n),
\end{equation}
where 
\begin{equation}
\mathbf{b}_j = \mathrm{vec}\Big( \Big\{ \begin{bmatrix} \mathbb{E}[X_iX_j] \\ \mathbb{E}[X_i]\end{bmatrix}, i\neq j \Big\}\Big).
\end{equation}

The subgradient $g(\theta)$ is computed by:
\begin{equation}
g(\theta) = \mathbf{E}[g(\theta;X)],
\end{equation} 
where $g(\theta;X)$ can be computed using \cref{algo:subgrad} as:
\begin{equation}
g(\theta;X) = \texttt{subgrad}(\theta;x),
\end{equation} 
substituting $\nabla_{\hat{x}}G_{j}(\hat{x};x)$ with:
\begin{equation}\label{eq:grad_general1}
\nabla_{\theta} G_j(\theta;x) = \mathrm{vec}( \mathbf{k}_1, \cdots, \mathbf{k}_{j-1}, \mathbf{0} , \mathbf{k}_{j+1}, \cdots, \mathbf{k}_{n}),
\end{equation}
where
\begin{equation}\label{eq:grad_general2}
\mathbf{k}_\ell = -2 \cdot \mathrm{vec}\Big( \Big\{ \begin{bmatrix}x_\ell (x_i - (w_{i\ell}x_\ell + b_{i\ell})) \\  (x_i - (w_{i\ell}x_\ell + b_{i\ell}))\end{bmatrix}, i\neq \ell \Big\}\Big).
\end{equation}
\bibliographystyle{IEEEtran}        
\bibliography{DDSS_draft.bib}

\end{document}